\documentclass[preprint,12pt]{elsarticle}
\usepackage{amssymb}
\usepackage{graphicx}

\journal{Physics Letters B}

\begin{document}

\title{
Constrained caloric curves and phase transition\\ for hot nuclei\\
INDRA Collaboration}

\author[ipno]{B.~Borderie}
\author[INFNFirenze]{S. Piantelli}
\author[ipno]{M.~F.~Rivet}
\author[ifin]{Ad. R. Raduta}
\author[ipno]{G.~Ademard}
\author[ganil]{E.~Bonnet}
\author[lpc]{R.~Bougault}
\author[ganil]{A.~Chbihi}
\author[ganil]{J.D.~Frankland}
\author[ipno,cnam]{E.~Galichet}
\author[ganil]{D.~Gruyer}
\author[lyon]{D.~Guinet}
\author[lyon]{P.~Lautesse}
\author[lpc]{N. Le Neindre}
\author[lpc]{O.~Lopez}
\author[ganil]{P.~Marini}
\author[lpc,ifin]{M.~P\^arlog}
\author[IFJ]{P.~Pawlowski}
\author[nap]{E.~Rosato}
\author[laval]{R.~Roy}
\author[nap]{M.~Vigilante}
\address[ipno]{Institut de Physique Nucl\'eaire, CNRS-IN2P3,
Universit\'e Paris-Sud 11, F-91406 Orsay Cedex, France}
\address[INFNFirenze]{INFN Sezione di Firenze, 50019  Sesto Fiorentino (FI), Italy}
\address[ifin]{National Institute for Physics and Nuclear Engineering,
RO-76900 Bucharest-Magurele, Romania}
\address[ganil]{GANIL, (DSM-CEA/CNRS-IN2P3), F-14076 Caen Cedex, France}
\address[lpc]{LPC Caen, ENSICAEN, Universit\'{e} de Caen, CNRS-IN2P3 
F-14050 Caen Cedex, France}
\address[cnam]{Conservatoire National des Arts et M\'etiers, F-75141,
Paris Cedex 03, France}
\address[lyon]{Universit\'e Claude Bernard Lyon 1,
Institut de Physique Nucl\'eaire, CNRS-IN2P3, F-69622 
Villeurbanne Cedex, France}
\address[IFJ]{Institute of Nuclear Physics, IFJ-PAN, 31-342 Krak\'ow, Poland}
\address[nap]{Dipartimento di Scienze Fisiche e Sezione INFN,
Universit\`a di  Napoli ``Federico II'', I-80126 Napoli, Italy}
\address[laval]{Universit\'e Laval, Qu\'ebec, G1V 0A6, Canada} 
\begin{abstract}
Simulations based on experimental data obtained from multifragmenting quasi-fused nuclei
produced in central $^{129}$Xe + $^{nat}$Sn collisions have been
used to deduce event by
event freeze-out properties in the thermal excitation energy range
4-12 AMeV [Nucl. Phys. A809 (2008) 111]. From these properties and the temperatures
deduced from proton
transverse momentum fluctuations, constrained caloric
curves have been built. At constant average volumes caloric curves
exhibit a monotonic behaviour whereas for constrained pressures
a backbending
is observed. Such results support the existence
of a first order phase transition for hot nuclei.
\end{abstract}

\begin{keyword}
Quasi-fusion reactions, nuclear multifragmentation, Caloric curves,
first order phase transition
\end{keyword}


\maketitle


One of the most important challenges of heavy-ion collisions at intermediate
energies is the identification and characterization of the nuclear
liquid-gas phase transition for hot nuclei, which has been theoretically
predicted for nuclear
matter~\cite{Sch82,Cur83,Jaq83,Mul95}. During the last fifteen years a big effort
to accumulate experimental indications of the phase transition has been
made. Statistical mechanics for finite systems appeared as a key issue to
progress, revealing new first-order phase transition signatures related
to thermodynamic anomalies like negative microcanonical heat capacity and
bimodality of an order parameter~\cite{Cho04,WCI06,Bor08,MDA00,I72-Bon09}.
Before this, correlated temperature and excitation energy measurements,
commonly termed caloric curves, were among the first possible
signatures to be studied~\cite{Bon85,Poc95,I8-Ma97,Nato02}.
However in spite of the observation
of a plateau in some caloric curves, no decisive conclusion related
to a phase transition
could be extracted~\cite{Dur98,Das01,I46-Bor02}. The reason is
that it is not possible to perform experiments
 at constant pressure or constant average
volume, which is required for an unambiguous phase transition signature.
Indeed, theoretical studies show that whereas many different caloric curves can
be generated depending on the path followed in the thermodynamical
landscape, constrained caloric curves must exhibit two
behaviours if a first order phase transition is present:
a monotonic evolution at constant average volume and a
backbending of curves at constant pressure~\cite{Cho00,Fur06}. 

In Ref.~\cite{I58-Pia05,I66-Pia08} we presented simulations
able to correctly reproduce
most of the experimental observables measured for
hot nuclei formed in central collisions (quasi-fused systems, QF, from
$^{129}Xe$+$^{nat}Sn$, 32-50 AMeV). 
The aim of the present Letter is to use  the event by event properties
at freeze-out which were inferred from these simulations to build constrained
caloric curves.

Experimental data were collected with the 4$\pi$
multidetector INDRA which is described in detail in Ref.~\cite{I3-Pou95,I5-Pou96}.
Accurate particle and fragment identifications were achieved
and the energy of the detected products was measured
with an accuracy of 4\%. Further details can be found 
in Ref.\cite{I14-Tab99,I33-Par02,I34-Par02}.
All the available experimental information
(charged particle energy spectra, average and standard deviation of fragment
velocity spectra and calorimetry) of selected QF
sources produced in central $^{129}$Xe+$^{nat}$Sn collisions which undergo
multifragmentation was used.

The method for reconstructing freeze-out
properties from simulations~\cite{I58-Pia05,I66-Pia08} 
 requires data with a very high degree
of completeness, 
 crucial for a good estimate of Coulomb energy.
QF sources are reconstructed, event by event,
from all the fragments and twice the charged particles emitted in the range
$60-120^{\circ}$ in the reaction centre of mass, in order to exclude
the major part of pre-equilibrium
emission~\cite{I29-Fra01,I69-Bon08}; with such a prescription only particles
with isotropic angular distributions and constant average kinetic energies are
considered. In simulations,
excited fragments and particles at
freeze-out are described by spheres at normal density.
Then the excited fragments subsequently deexcite while flying apart.
Four free parameters
are adjusted to fit the data at each incident energy: the percentage
of measured particles which were
evaporated from primary fragments, the collective radial energy, a minimum
distance between the surfaces of products at freeze-out and a limiting
temperature for fragments. All the details of simulations can be found
in Ref.~\cite{I58-Pia05,I66-Pia08}.
The limiting temperature,
related to the vanishing of level density for fragments~\cite{Koo87},
was mandatory to reproduce the observed widths of fragment velocity
spectra.
Indeed, the sum of Coulomb repulsion, collective energy, 
thermal kinetic energy 
and spreading due to fragment
decays accounts only for about 60-70\% of those widths.
By introducing a limiting temperature for fragments, the thermal kinetic
energy increases, due to energy conservation, which produces the missing
percentage for the widths of final velocity distributions.
The agreement between experimental and simulated velocity/energy spectra for
fragments, for the
different beam energies, is quite remarkable 
(see figure 3 of~\cite{I66-Pia08}).
Relative velocities between fragment pairs were also compared
through reduced relative velocity correlation 
functions~\cite{Kim92,I57-Tab05}
(see figure 4 of~\cite{I66-Pia08}).
Again a good agreement is obtained between experimental data and
simulations, which
indicates that the retained method (freeze-out topology built up
at random) and the deduced parameters are sufficiently relevant
to correctly describe the freeze-out configurations, including volumes.
However it should be noted that the agreement between experimental and simulated 
energy spectra for protons and
alpha-particles (see figure 5 of~\cite{I66-Pia08}) is not so good; this
may come from the fact that we have chosen a single value, at each
incident energy, for the percentage of all measured
particles which were evaporated from primary fragments
to limit the number of parameters of
the simulation. We shall come back to this point later. 

From the simulations we deduce, event by event, various
quantities needed to build constrained caloric curves, namely
the thermal excitation energy of QF hot nuclei, $E^*$ (total excitation minus
collective energy) with an estimated systematic error of around 1 AMeV,
the freeze-out volume $V$ (see envelopes of figure 8 from~\cite{I66-Pia08}) and
the total thermal kinetic energy at freeze-out $K$.
Events are sorted into $E^*$ bins of 0.5 AMeV with their
associated kinetic temperature $T_{kin}$ at freeze-out.
In simulations, Maxwell-Boltzmann statistics
are used for particle velocity distributions at freeze-out
and consequently
the deduced temperatures, $T_{kin}$, are classical.
It is important to stress here that, at present time,
there is no unique thermometer and, depending
on the excitation energy range, disagreements can be observed between kinetic,
chemical temperatures and temperatures deduced from excited
states~\cite{Das01,I46-Bor02,Ser98,Tra98}.

With regard to the pressure at freeze-out, it can be
derived within the microcanonical ensemble.
Let us consider fragments interacting 
only by Coulomb and excluded volume,
which corresponds to the freeze-out configuration.
Within a microcanonical ensemble, the statistical weight of a configuration
$C$, defined by the mass, charge and internal excitation energy
of each of the constituting $M_C$ fragments,
can be written as
\begin{eqnarray}
\nonumber
W_C(A,Z,E,V) = \frac1{M_C!} \chi V^{M_C} \prod_{n=1}^{M_C}\left( 
\frac{\rho_n(\epsilon_n)}{h^3}(mA_n)^{3/2}\right)
\\ 
\times
~ \frac{2\pi}{\Gamma(3/2(M_C-2))} ~ \frac{1}{\sqrt{({\rm det} I})}
~ \frac{(2 \pi K)^{3/2M_C-4}}{(mA)^{3/2}},
\label{eq:wc}
\end{eqnarray}
where $A$, $Z$, $E$ and $V$ are respectively the mass number,
the atomic number, the excitation energy and
the freeze-out volume of the system.
$E$ is used up in fragment formation, fragment internal
excitation, fragment-fragment Coulomb interaction and
thermal kinetic energy $K$.
$I$ is the inertial tensor of the system whereas
$\chi V^{M_C}$ stands for the free volume or, equivalently, accounts for
inter-fragment interaction in the hard-core idealization. 

The microcanonical equations of state are
\begin{eqnarray}
\nonumber
T=\left(\frac{\partial S}{\partial E}\right)^{-1}|_{V,A},\\
\nonumber
P/T=\left(\frac{\partial S}{\partial V}\right)|_{E,A},\\
-\mu/T=\left(\frac{\partial S}{\partial A}\right)|_{E,V}.
\end{eqnarray}

Taking into account that $S=\ln Z=\ln \sum_C W_C$ 
and that $\partial W_C/\partial V=\left(M_C/V\right) W_C$, 
it comes out that
\begin{eqnarray}
\nonumber
P/T=
\left(\frac{\partial S}{\partial V}\right)&=&\frac1{\sum_C W_C} \sum_C 
\frac{\partial W_C}{\partial V}\\
&=&\frac1V \frac{\sum_C M_C W_C}{\sum_C  W_C}=\frac{<M_C>}{V}.
\end{eqnarray}

The microcanonical temperature is also easily deduced from its statistical
definition~\cite{Radut02}:
\begin{eqnarray}
\nonumber
T=\left(\frac{\partial S}{\partial E}\right)^{-1}&=&(\frac1{\sum_C W_C} \sum_C 
W_C(3/2M_C-5/2)/K)^{-1}\\
&=&<(3/2M_C-5/2)/K>^{-1}.
\end{eqnarray}

As $M_C$, the total multiplicity at freeze-out, is large, we have
\begin{eqnarray}
T\thickapprox
\frac{2}{3}<\frac{K}{M_C}>
\end{eqnarray}

and the pressure $P$ can
be approximated by
\begin{eqnarray}
P=T\frac{<M_C>}{V}\thickapprox\frac{2}{3}\frac{<K>}{V}.
\end{eqnarray}
Knowing $<K>$ and $V$ from simulations, pressure $P$ can be calculated for
events sorted in each $E^*$ bin. The temperature $T_{kin}$ that we obtain from
the simulations is identical to the microcanonical temperature of
equation (5). One can also note that the free Fermi gas pressure
exactly satisfies equation (6). 

\begin{figure}
\begin{center}
\includegraphics*[width=0.48\textwidth]
{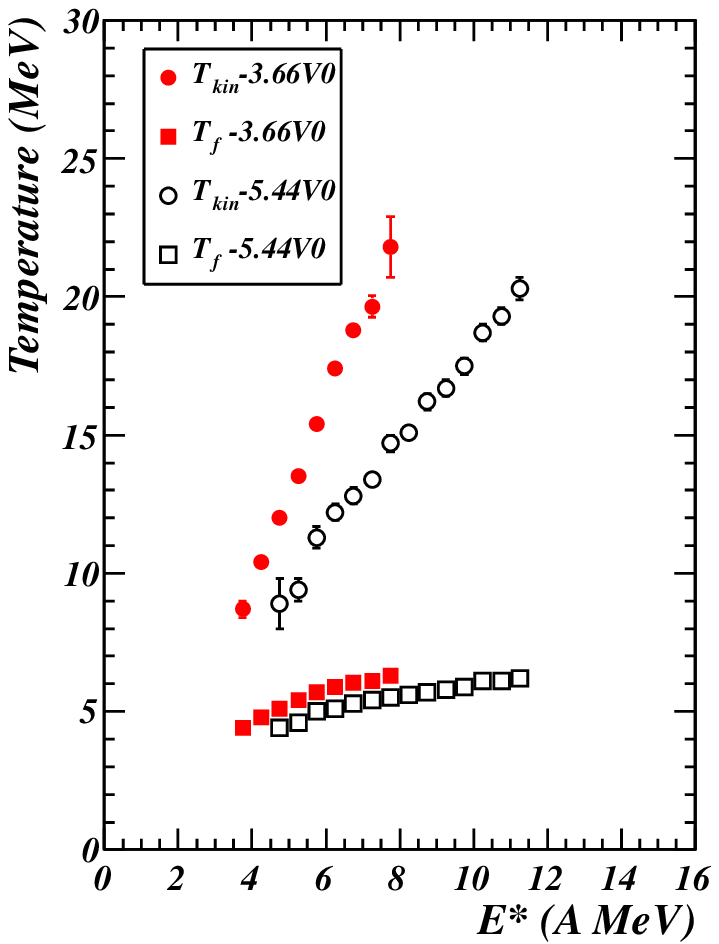}
\includegraphics*[width=0.48\textwidth]
{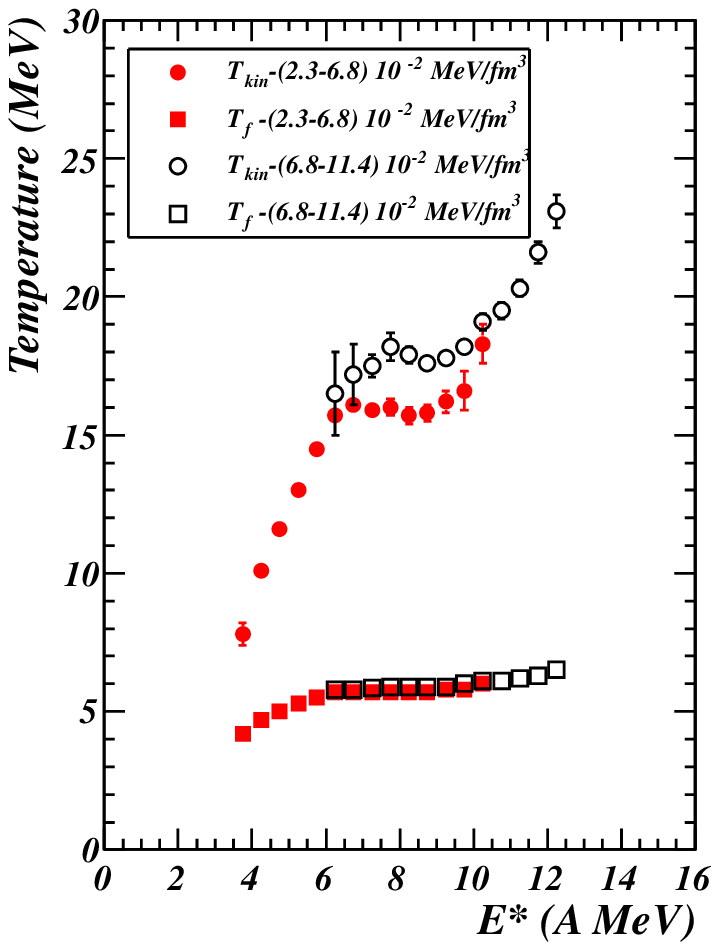}
\end{center}
\caption{(Color online) Caloric curves (kinetic temperature $T_{kin}$ versus thermal
excitation energy $E^*$) constrained at average volumes (left)
and for selected ranges of pressure (right) and the corresponding internal
temperatures of fragments $T_f$. Error bars correspond to statistical errors.}
\label{fig1}
\end{figure}

Constrained caloric curves, built with correlated values of
$E^*$ and $T_{kin}$ have been derived for QF hot nuclei with $Z$ restricted
to the range 80-100, which corresponds to the $A$ domain 194-238, in order to
reduce effects of mass variation on caloric curves~\cite{Nato02};
they are presented in Fig.~\ref{fig1}.
Curves for internal fragment temperatures, $T_f$, are also shown in the figure.
For two different average freeze-out volumes
 corresponding to the ranges
3.0-4.0$V_0$ and 5.0-6.0$V_0$
- where $V_0$ is
the volume of the QF nuclei at normal density -
a monotonic behaviour of caloric curves is observed
as theoretically expected.
The caloric curves when pressure ranges have been selected 
exhibit a backbending and moreover their qualitative evolution 
with increasing pressure exactly corresponds to what is theoretically
predicted with a microcanonical lattice gas model~\cite{Cho00}.
The decrease of $T_{kin}$ occurs in the $E^*$ region
where $<M_C>$/$V$ increases faster with $E^*$ than in the surrounding
regions, in agreement with expected
spinodal fluctuations~\cite{Cho04}.

By extrapolating to higher pressures, 
one could infer a critical temperature -vanishing of backbending -
around 20 MeV. Such a value is
within the range calculated for infinite nuclear matter
whereas a lower value is expected for finite systems 
in relation with surface
and Coulomb effects~(see~\cite{I46-Bor02} and references therein).
We thus wonder if the classical
temperature $T_{kin}$ is relevant.

Very recently a new method for measuring the temperature of hot nuclei was
proposed~\cite{Wu10,Zhe11}. It is based on momentum fluctuations of emitted
particles, like protons, in the centre of mass frame of the fragmenting nuclei.
On the
classical side, assuming a Maxwell-Boltzmann distribution of the momentum
yields, the temperature $T$ is deduced from the quadrupole momentum fluctuations
defined in a direction transverse to the beam axis:\\
$\sigma^2$ = $<Q_{xy}^2>$ - $<Q_{xy}>^2$ = $4m^2T^2$\\
with $Q_{xy}$ = $p_{x}^2$ - $p_{y}^2$;
$m$ and $p$ are the mass and linear momentum of emitted particles.
Taking into account the quantum nature of particles, a correction
$F_{QC}$
related to a Fermi-Dirac distribution was also proposed~\cite{Zhe11,Zhe12}.\\
In that case $\sigma^2$ = $4m^2T^2$ $F_{QC}$ where $F_{QC}$ =
$0.2(T/\epsilon_f)^{-1.71}$ + 1;\\ $\epsilon_f$ = 36 $(\rho/\rho_0)^{2/3}$
is the Fermi energy of nuclear matter at density $\rho$ and
$\rho_0$ corresponds to normal density. 

\begin{figure}
\begin{minipage} {17pc}
\vspace{1.0pc}
\includegraphics[width=17pc]{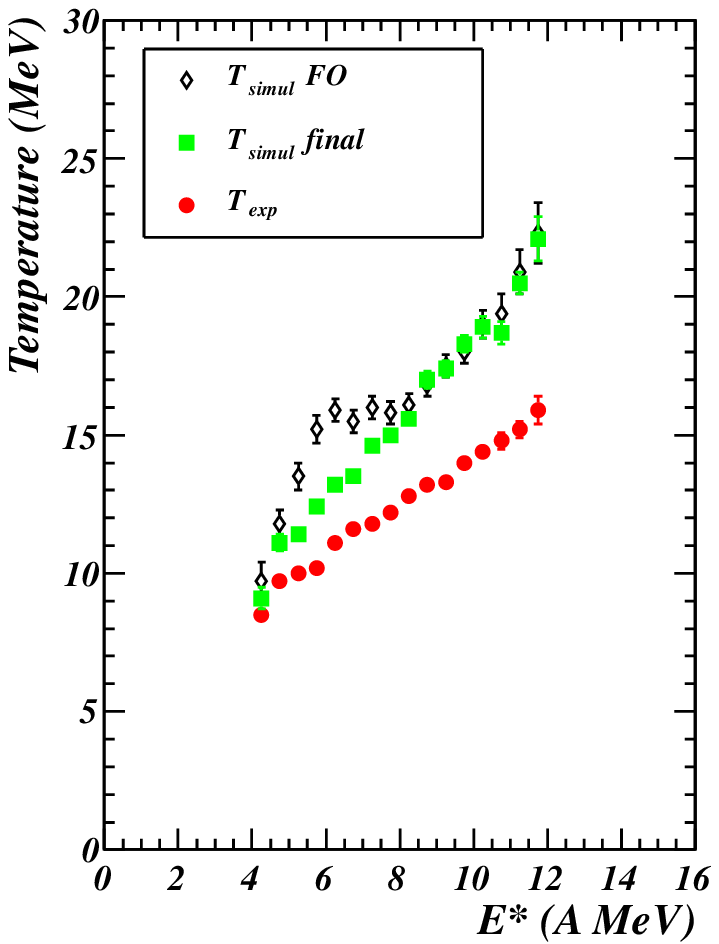}
\caption{\label{fig2}
(Color online) Caloric curves (classical temperature from proton transverse momentum
fluctuations versus thermal excitation energy) for protons (simulation)
thermally emitted at freeze-out (open diamonds), for protons (simulation)
after the secondary decay stage (full squares),
and from protons experimentally measured (full points). Error bars
correspond to statistical errors}
\end{minipage}\hspace{2pc}%
\begin{minipage} {17pc}
\includegraphics[width=17pc]{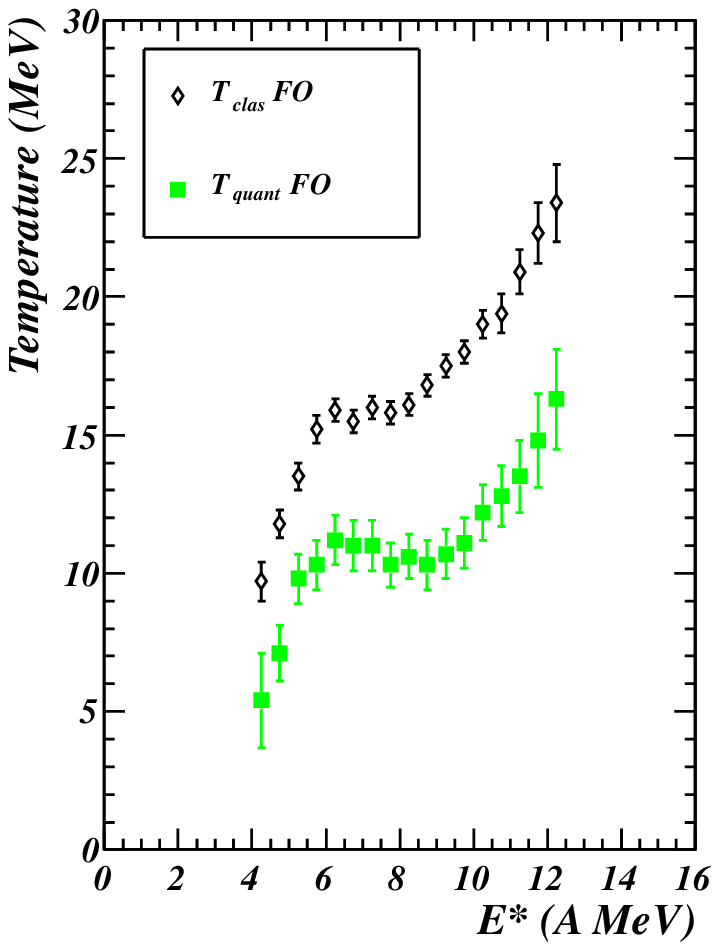}
\caption{\label{fig3} (Color online) Caloric curves: classical temperature (open diamonds)/
quantum corrected temperature (full squares) 
from proton transverse momentum
fluctuations versus thermal excitation energy. Protons (simulation)
are thermally emitted at freeze-out. Error bars include statistical and
systematic errors.}
\end{minipage}
\end{figure}

Before using this new thermometer (with protons) to build constrained
caloric curves, it was
important to have made several verifications. With the classical simulation
(freeze-out and asymptotic proton momenta), it
is possible to test the agreement with the proposed classical thermometer.
Moreover the effects of secondary decays on temperature measurements can be
estimated. Fig.~\ref{fig2} shows different caloric curves without constraints.
Note that the selection in $Z$ and $A$ of hot nuclei is the same as in the
previous figure; it was also verified that, within statistical errors,
at a given thermal excitation energy, transverse momentum fluctuation values
are the same for our selection or
by selecting only a single ($A$ and $Z$) hot nucleus. 
Open diamonds refer to classical temperatures calculated from momentum fluctuations
for protons thermally emitted at freeze-out. Within statistical errors
they perfectly superimpose on unconstrained $T_{kin}$ values.
Full squares correspond to classical temperatures calculated from
momentum fluctuations for protons after the secondary decay stage. We note
that the caloric curve is distorted, which means that it is hazardous to
use experimental data from protons to measure temperatures. Moreover, in
this case, quantum corrections for temperatures can not be made because protons
are emitted at different stages of deexcitation with different Fermi energy values.
In Fig.~\ref{fig2} classical temperatures calculated from experimental proton
data are also shown (full points). As for temperatures calculated from asymptotic proton
data of simulations, a monotonic behaviour of the caloric curve is observed.
One also notes the differences between the two sets of temperature values,
which are related to the
fact that, as indicated previously, simulations do not describe
accurately the experimental proton energy spectra. For each $E^*$ value
the difference $\Delta T$= $T_{simul}$ - $T_{exp}$ between final temperature
from proton data from simulations and temperature from experimental protons
will be used to correct classical temperatures derived from simulated protons
at freeze-out.

It finally appears that the only way to extract temperatures from proton transverse momentum
fluctuations taking into account quantum effects is to use protons thermally
emitted at freeze-out. In that case classical temperature values from simulations
must be extracted and corrected and then, quantum corrections applied, which
needs Fermi energy values. Those values can be estimated from semi-classical
calculations (Xe+Sn at 32 AMeV and Sn+Sn at 50 AMeV)~\cite{I29-Fra01,Riz07}:
protons thermally emitted at freeze-out at time around 100-120 fm/c after the
beginning of collisions come from a low density uniform source. For the two
incident energies low densities around $\rho\sim0.4\rho_0$ are calculated,
which corresponds to $\epsilon_f\sim$ 20 MeV. We have introduced a systematic
error of $\pm$ 0.1$\rho_0$ for the calculation of $\epsilon_f$ and consequently
a systematic error for ``quantum'' temperatures of $\pm$ 0.6 to $\pm$ 0.5 MeV on the
considered temperature range. Fig.~\ref{fig3} shows the final
caloric curve with temperatures from quantum fluctuations (full squares).
It exhibits a plateau around a temperature of 10-11 MeV on the $E^*$ range
5-10 AMeV.
For comparison the caloric curve with classical temperatures derived from
the simulation and presented in Fig.~\ref{fig2} is added (open diamonds).
\begin{figure}
\begin{center}
\includegraphics*[width=0.48\textwidth]
{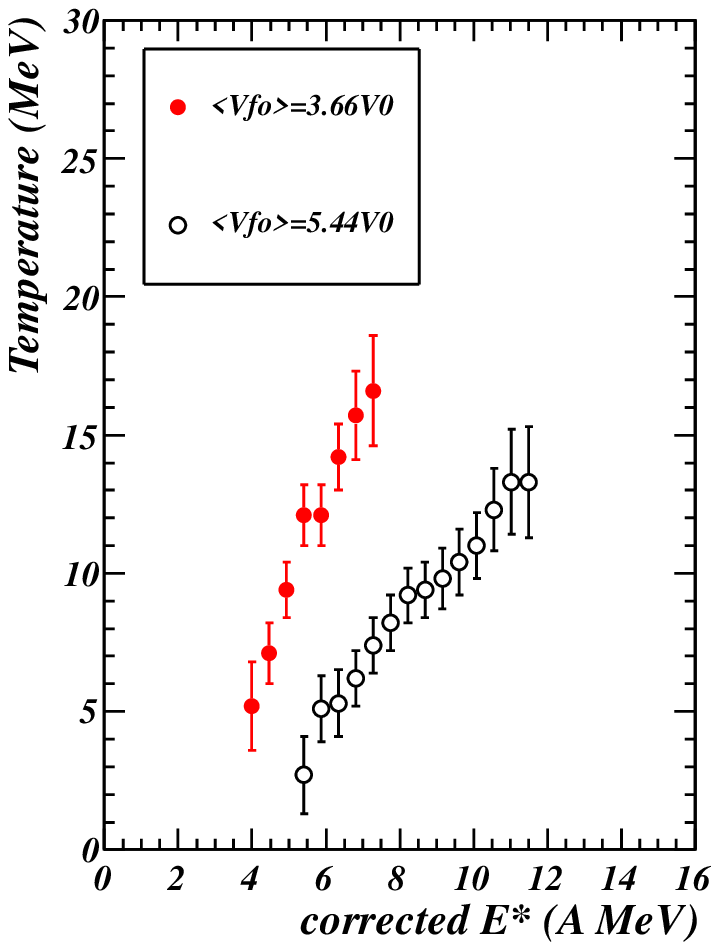}
\includegraphics*[width=0.48\textwidth]
{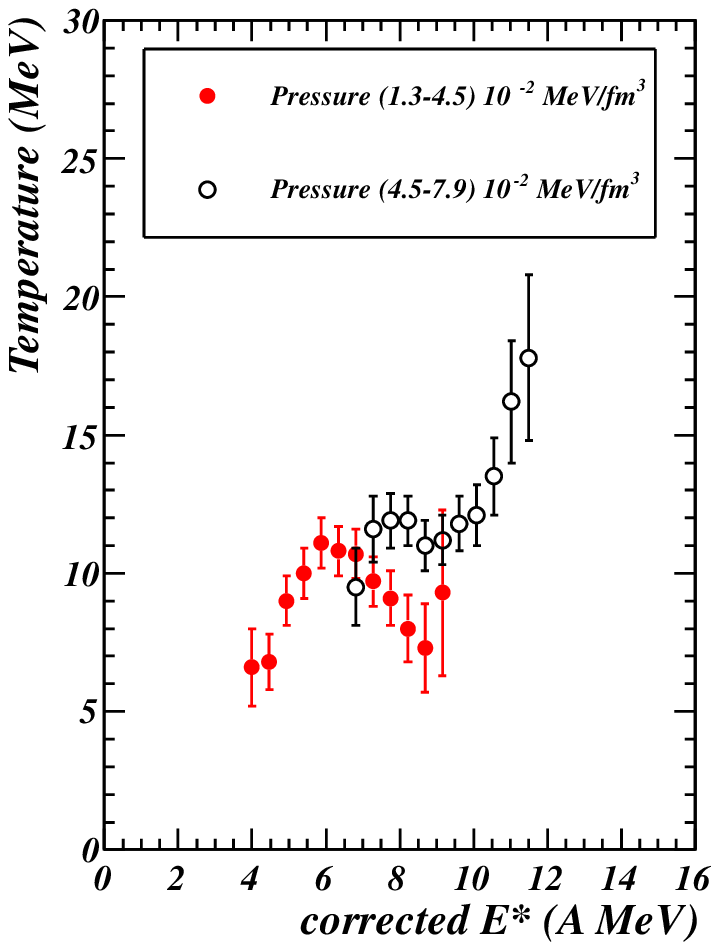}
\end{center}
\caption{(Color online) Caloric curves (quantum corrected temperature versus
corrected thermal
excitation energy) constrained at average volumes (left)
and for selected ranges of pressure (right).
Error bars include statistical and
systematic errors.}
\label{fig4}
\end{figure}

Constrained caloric curves, which correspond to correlated values of 
$E^*$ and quantum corrected temperatures have been determined.
The $E^*$ values have been corrected \textit{a posteriori}.
Indeed they are derived from experimental
calorimetry plus estimated kinetic energy for neutrons emitted at freeze-out
($E_n^{fo}$ = $M_n^{fo}\times3T/2$ - see~\cite{I58-Pia05}),
which has been modified using quantum temperatures instead of classical ones.
Pressure values were also corrected
using quantum temperatures in equation (6).
In Fig.~\ref{fig4} (left) we have constructed caloric curves for
the two 
average freeze-out volumes previously chosen.
Again as theoretically expected a monotonic behaviour of caloric curves is observed.
Fig.~\ref{fig4} (right) shows the caloric curves when pressure has been constrained
within two domains: (1.3-4.5) and (4.5-7.9)x $10^{-2}$ MeV fm$^{-3}$.
Backbending is seen especially for the lower pressure range. 
For higher pressures the backbending of the caloric curve is reduced
and one can estimate its vanishing, indicating the critical temperature,
around 13 MeV for
the selected finite systems. Moreover,
we can also estimate the upper limits of the spinodal
region and of the coexistence region (see Fig. 2.1 of~\cite{Cho04})
around respectively 8 and 10 AMeV.
Those estimates are
in good agreement (within error bars) with spinodal~\cite{I40-Tab03} and
bimodality~\cite{I72-Bon09} signals. 

\begin{figure}
\begin{center}
\includegraphics*[width=0.48\textwidth]
{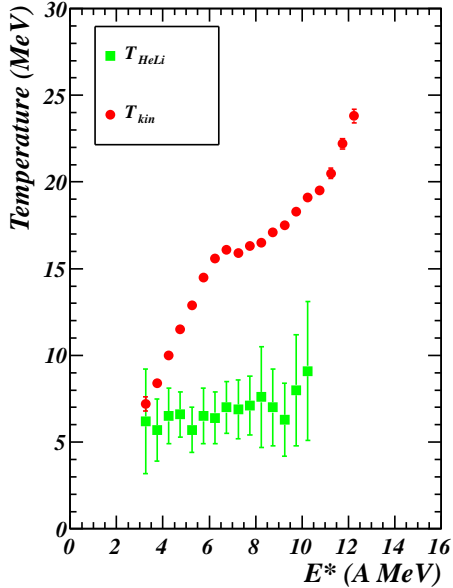}
\end{center}
\caption{(Color online) Caloric curves obtained with
two different temperature measurements: $T_{kin}$ and $T_{HeLi}$.
$T_{kin}$ values are obtained from the simulation whereas
$T_{HeLi}$ values
are derived from experimental data.
Error bars correspond to statistical errors.}
\label{fig5}
\end{figure}

As far as internal temperature of fragments are concerned (see
Fig.~\ref{fig1}), one observes that the values from
the simulation, $T_f$,
perfectly agree, for our range of mass 194-238,
with those calculated with the well known ``He/Li thermometer''~\cite{Poc95,Nato02}.
We also apply  this thermometer, keeping the prefactor 16
proposed in Ref.~\cite{Poc95}, to the experimental data. The
derived temperature values are presented in
Fig.~\ref{fig5} with the $T_{kin}$ values from the simulations.
We observe that the measured $T_{HeLi}$ also exhibit a plateau
and are close to the fragment temperatures, $T_f$, of Fig.~\ref{fig1}.
Note that, in the excitation energy range 4-10 AMeV, temperatures
extracted from $^{5}Li$ excited states also agree
with $T_{HeLi}$~\cite{I46-Bor02,Tra98}. 
This is an indication that the temperatures obtained with the He/Li thermometer 
seem to reflect the internal temperature of fragments in this excitation
energy range.
In Ref.~\cite{Nato02} the evolution with the mass of hot nuclei of those plateau
temperatures is assimilated to that
of limiting temperatures
resulting from Coulomb instabilities
of heated nuclei predicted long ago~\cite{Lev85}.
The following explanation can be given.
For thermally equilibrated QF hot nuclei one expects internal temperature
for simultaneously emitted fragments equal to the temperature of the fragmenting
system, which can not exceed its Coulomb-related limiting temperature.
As a direct consequence the
internal fragment temperatures must reflect the evolution of this limiting
temperature with the mass of hot nuclei. 
Such an explanation is supported by two experimental results: on one
side, the fact that, on average, thermal equilibrium is achieved
at the freeze-out stage~\cite{I11-Mar98} and, on the other side, the observation of a
limitation of excitation energy for fragments on the considered $E^*$
range~\cite{I39-Hud03}.

In conclusion, several caloric curves have been derived for quasi-fused
systems using a new thermometer based on proton transverse momentum fluctuations
including quantum effects. The unconstrained caloric curve exhibits a plateau at
a temperature around 10-11 MeV on the thermal excitation energy range
5-10 AMeV. For constrained caloric curves (volume or pressure) we observe
what is expected for a first order phase transition for finite systems in
the microcanonical ensemble, namely a monotonic behaviour at constant
average volume and backbending for constrained pressure.
After the observation of negative microcanonical heat capacity and bimodality
of the heaviest fragment distribution, this behaviour of caloric curves
is the ultimate signature of a
first order phase transition for hot nuclei. 

The only piece now missing is the nature of the dynamics of
the transition, i.e. the fragment formation. Two mechanisms have
been proposed. On one side, stochastic mean field approaches for which the
fragmentation process follows the spinodal fragmentation scenario and, on the
other side, molecular dynamics (QMD, AMD) for which many-body correlations play a
stronger role and pre-fragment appear at earlier
times~\cite{Cho04,Har98,Ono99,Riz07,Lef09}. From the experimental
side, signals in favor of spinodal fragmentation were observed but confidence levels
around 3-4 $\sigma$ prevent any definitive
conclusion~\cite{I31-Bor01,I40-Tab03}. Analyses of new experiments with higher
statistics, are in progress.




\end{document}